\documentstyle[12pt]{article}
\global\arraycolsep=2pt

\newcommand{\k}{\vec{k}_{\perp}^2}
\newcommand{\be}{\begin{eqnarray}}
\newcommand{\ee}{\end{eqnarray}}
\newcommand{\la}{\langle}
\newcommand{\ra}{\rangle}

\newcommand{\Dm}{\vec{iD}_{\mu} }
\begin{document}

\begin{titlepage}
\begin{flushright}
SMU-HEP-93-25\\
hep-ph 9401278\\
December 1993
\end{flushright}

\vspace{0.3cm}
\begin{center}
\Large\bf The Nonperturbative Wave Functions,\\ Transverse Momentum
Distribution  \\
 and\\
QCD Vacuum Structure.
 \end{center}

\vspace {0.3cm}

 \begin{center} {\bf Ariel R. Zhitnitsky\footnote{
On leave of absence from Budker Institute of Nuclear Physics,\\
Novosibirsk,630090,Russia.\\
e-mail addresses:arz@mail.physics.smu.edu, ariel@sscvx1.ssc.gov}}
 \end{center}

\begin{center}
{\it Physics Department, SMU , Dallas, Texas, 75275-0175}

\end{center}
\begin{abstract}

It is shown that there is one-to-one correspondence between
two, apparently different problems: \\
1. The determination of the meanvalues of transverse moments
 $\la \vec{k}_{\perp}^{2n} \ra $  for the  nonperturbative
pion wave function $\psi(\k, x)$ and \\
2. The evaluation of the mixed vacuum condensates
 $ \la \bar{q}G_{\mu\nu}^n
q\ra $.

Arguments in favor of a large magnitude   of the mixed
vacuum condensate
$ \la \bar{q}(ig\sigma_{\mu\nu} G_{\mu\nu})^2 q \ra$  are given.
The analysis is
based on   the dispersion relations and  PCAC.
Because of the large values of the
condensates we found a noticeable fluctuations of the momentum
$\la \vec{k}_{\perp}^4 \ra  > \la \k\ra^2$.
 We also found  some general
 properties
of the condensates
$ \la \bar{q}(ig\sigma_{\mu\nu} G_{\mu\nu})^nq \ra$ for arbitrary $n$.
This information is used for the analysis of the higher
 moments $\la \vec{k}_{\perp}^{2n} \ra $
 in the limit when the space-time dimension $d\rightarrow\infty$.
As a byproduct,
it is proven that the  standard assumption
on factorizability of the $\psi(\k, x ) =\psi(\k )\phi(x)$
{\bf does contradict} to the very general properties of the
theory.
  We define and model $\psi(\k, x)$, satisfying all these constraints.

\end{abstract}
\end{titlepage}
\vskip 0.3cm
\noindent
{\bf 1. Introduction}

 The problem of bound states in the
relativistic quantum field theory with
large coupling constant is an extremely difficult
problem. To  understand the structure of the
bound state is a very ambitious goal which
assumes the
solution of a whole spectrum of tightly connected problems, such as
confinement,
chiral symmetry breaking phenomenon,   and many, many others which
are greatly important in the low energy region.

A less ambitious purpose is the  study of
 the hadron wave function ($wf$) with a minimal number
of constituents.
As is known such a function gives  parametrically leading contributions
to hard exclusive processes. In this case the quark and antiquark are
produced at small distances $z\sim 1/Q\rightarrow 0$, where $Q$
is the typical large momentum transfer. Thus, we can neglect
the $z^2$ dependence of the wave function of the
meson with momentum $p$ and can concentrate on the
variable  $zp \simeq 1$ which is order of one. Therefore,
the problem is
drastically simplified in the asymptotic limit and we end up
with the light cone wave function, $\phi(zp, z^2=0) $.

The corresponding
wave functions have  been introduced to the theory
 in the late seventies
and  early eighties  \cite{Brod} in order to describe the
exclusive processes in QCD. We refer to the review papers
\cite{Cher},\cite{Brod1}, \cite{Cher1}
 on this subject for the  detail definitions
and discussions in the given context, but here we want to make
a short remark concerning the very unusual properties of the
nonperturbative light-cone wave functions of
the leading twist. The information which can be
extracted from  the QCD sum rules method \cite{Shif1},\cite{Shif2}
unambiguously shows the asymmetric form of the distribution amplitudes
and this property was unexpected and even suspicious to many physicists.

However, to  study   the fine aspects of the
 theory (see explanations below)
of the exclusive processes and to extend the area of applications,
we need  to know not only the dependence $wf$ on the longitudinal
variable $x_i$, but on  transverse variable $\k$ as well.
In particular, as is known \cite{Ster},\cite{Kroll}, the Sudakov
suppression
should be taken into account in order to  integrate correctly
over the  `` endpoint region" ($x_i\rightarrow 1$). The
dependence of $wf$
on  the transverse momentum $\k$ plays an important role in
such calculation. Besides that, the transverse size
 dependence plays a
key role in color transparency physics.

$\bullet$ The main goal of the paper is the calculation
of the few lowest moments $\la \vec{k}_{\perp}^{2n} \ra $
 of the transverse
distribution.
Besides that we formulate
the definition of nonperturbative wave function and find
some  constraints  on it. In particular, we analyze
the asymptotically distant
terms $\la \vec{k}_{\perp}^{2n} \ra ,~~n\rightarrow\infty$
in order to model the $\psi (\k ,x_i)$. The byproduct
of this consideration is analysis of
the factorizability of  $\psi (\k ,x_i)=\phi(x_i)\psi(\k)$, which
it turns out, does not work. Finally, we model $wf$ which meets all
constraints.
Analogous problem for the longitudinal distribution
$\la \xi^n \ra$  has been formulated and solved within
QCD sum rules method in ref.\cite{Cher2}.

The paper is organized as follows.
In the next section we derive the explicit relations between
two, apparently different, values:\\
1.$\la \vec{k}_{\perp}^{2n} \ra_P $  for pseudoscalar $wf$, $\phi_P=
\la 0|\bar{d}(z)i\gamma_5u(-z)|\pi\ra $ on the one side and \\
2. vacuum expectation values (VEVs)
$ \la \bar{q}G_{\mu\nu}^n q\ra $ on the other side.
Such exact (in chiral limit)
connection is not very surprising because $\pi$ meson
is the Goldstone particle, strongly interacting with
vacuum fluctuations. Well-known example of such connection is,
of course,
the formula
\be
\label{1}
\la 0|\bar{d} i\gamma_5u |\pi\ra=-\frac{1}{f_{\pi}}
\la 0|\bar{d}d +\bar{u}u |0\ra ,
\ee
which relates $\pi$ meson matrix element and some vacuum
characteristic.

In the section 3 we derive the analogous relations for
 $\la \vec{k}_{\perp}^{2n} \ra_A $  for axial (leading twist) $wf$.
In this case
the corresponding formulae are not exact even in the chiral limit,
but we argue that their accuracy is very high and corrections
are of order $\alpha_s/\pi\simeq 10\%$. We will see
that for both cases
(axial and pseudoscalar ) the problem
is reduced to the evaluation of the mixed vacuum condensates
$ \la \bar{q}G_{\mu\nu}^n q\ra $.

 In order to estimate them we consider in section 4 some special
kind of sum rules which are very sensitive to the VEVs we are
interested in. We find a large violation ( order of factor 3)
of factorization for mixed condensates $ \la \bar{q}
(\sigma_{\mu\nu}G_{\mu\nu})^2 q\ra $
of dimension seven. It automatically leads to a noticeable
fluctuations of the momentum $\la \vec{k}_{\perp}^{4} \ra_{A}
 >\la \k\ra_{A}^2$
 and leads to the well-spread out $wf$ (in the momentum
space). Let us note, that if the factorization
would work, we would get $\la \vec{k}_{\perp}^{4} \ra_{A,P}
 \approx \la \k\ra_{P,A}^2$. Such a relation gives the
{\it qualitatively} different behavior for $wf$ and means
a strong concentration   of the distribution density around
point $\k=0$.

In the section  we derive an ``almost" exact relation
between mixed vacuum
condensates $\la\bar{q}(ig\sigma_{\lambda\sigma}
 G_{\lambda\sigma})^n q\ra=(m_0^2)^n\la \bar{q}q\ra $
for arbitrary $n$.

The section 5 is the {\bf main part }of this paper.
We formulate the definition of
nonperturbative wave function and give some constraints on it.
Finally we model it.
 \vskip 0.3cm
\noindent
{\bf 2. The   QCD vacuum condensates and
 $\la \vec{k}_{\perp}^{2n} \ra_P$.
  }

We define the pion
pseudoscalar wave function $\la 0|\bar{d}(z)i\gamma_5
 u(-z)|\pi(q)\ra
=\frac{-2\la \bar{q}q\ra}
{f_{\pi}}\phi_P$ and the corresponding
mean values of the quark transverse distribution
in the following way:
\be
\label{2}
\la 0|\bar{d}i\gamma_5 (\Dm t_{\mu})^{2n}  u|\pi(q)\ra=
-\frac{2\la \bar{q}q\ra}
{f_{\pi}}(-t^2)^n\frac{(2n-1)!!}{(2n)!!}\la \vec{k}_{\perp}^{2n} \ra_P ,
\ee
where $\Dm=i\vec{\partial_{\mu}}+gA_{\mu}^a\frac{\lambda^a}{2}$ is the
covariant derivative and  a transverse vector $t_{\mu}$ is perpendicular
 to the hadron momentum $q_{\mu}$. The factor $\frac{(2n-1)!!}{(2n)!!}$
is related to the integration over $\phi$ angle in the transverse plane:
$\int d\phi (\cos\phi)^{2n}/ \int d\phi= {(2n-1)!!}/{(2n)!!}$.

We removed the common factor $-\frac{2\la \bar{q}q\ra}
{f_{\pi}}$ in order to reproduce the matrix element (\ref{1}) without
derivatives. To find $\la \vec{k}_{\perp}^{2} \ra_P$ we follow the paper
\cite{Zhit2} and consider the
following matrix element
\be
\label{3}
\la 0|\bar{d}i\gamma_5 \Dm\vec{iD}_{\nu} u|\pi(q)\ra=
-\frac{2\la \bar{q}q\ra}
{f_{\pi}}Ag_{\mu \nu},~~ \la \vec{k}_{\perp}^{2} \ra_P=-2A,
\ee
where we dropped off the kinematical structure
$q_{\mu}q_{\nu}$ by the
reason which will be clear soon.
Multiplying (\ref{3}) by $g_{\mu\nu}$
 and using the equations of  motion  and PCAC,
we get (in the chiral limit, $m_q\rightarrow 0$ ):
\be
\label{4}
\la 0| \bar{d}i\gamma_5 \Dm\Dm u|\pi(q)\ra=
-\frac{2\la \bar{q}q\ra}
{f_{\pi}}4A=\frac{1}{f_{\pi}}
\la \bar{q}ig\sigma_{\mu\nu}
G_{\mu\nu}^a\frac{\lambda^a}{2} q \ra,\\   \nonumber
 \la \vec{k}_{\perp}^{2} \ra_P \simeq
\frac{\la\bar{q}ig\sigma_{\mu\nu}
 G_{\mu\nu}^a\frac{\lambda^a}{2} q \ra }{4\la \bar{q}q\ra}
\simeq \frac{ m_0^2}{4}\simeq 0.2 GeV^2, ~~~m_0^2\simeq 0.8 GeV^2.
\ee
where we use the standard value for parameter $m_0^2$
\cite{Shif2}.
It is clear that the skipped term is proportional to
$q_{\mu}q_{\nu}g_{\mu\nu}\sim m_{\pi}^2\sim m_q$
and gives zero contribution in the chiral limit.

In order to find $\la \vec{k}_{\perp}^{4} \ra_P$ we
consider the following
matrix element
\be
\label{6}
\la 0|\bar{d}i\gamma_5\vec{iD}_{\mu_1
}\vec{iD}_{\mu_2}\vec{iD}_{\mu_3}\vec{iD}_{\mu_4}
 u|\pi(q)\ra=-\frac{2\la \bar{q}q\ra}
{f_{\pi}}\\    \nonumber
[Ag_{\mu_1\mu_2}g_{\mu_3\mu_4}+
Bg_{\mu_1\mu_3}g_{\mu_2\mu_4}+Cg_{\mu_1\mu_4}g_{\mu_3\mu_2}],
\ee
where we dropped off the terms which include the
$q_{\mu}q_{\nu}$ structure by the reasons mentioned above. The PCAC
leads to the following relation
\be
\label{7}
\la 0|\bar{d}i\gamma_5\vec{iD}_{\mu_1
}\vec{iD}_{\mu_2}\vec{iD}_{\mu_3}\vec{iD}_{\mu_4}
 u|\pi(q)\ra=-\frac{2 }{f_{\pi}}\la\bar{q}P_{\mu_1}P_{\mu_2}P_{\mu_3}
P_{\mu_4}q\ra,
\ee
where $P_{\mu }=\Dm $ is hermitian operator and
  condensate $\la\bar{d}P_{\mu_1}P_{\mu_2}P_{\mu_3}
P_{\mu_4}q\ra$ can be evaluated in a standard way. The
result is:
\be
\label{8}
\la \vec{k}_{\perp}^{4} \ra_P=
\frac{-\la \bar{q}g^2\sigma_{\mu\nu}
 G_{\mu\nu} \sigma_{\lambda\sigma}
 G_{\lambda\sigma}q \ra+2\la \bar{q}g^2
 G_{\mu\nu}   G_{\mu\nu}q \ra}{12\la\bar{q}q\ra},
\ee
where $G_{\mu\nu}=G_{\mu\nu}^a\frac{\lambda^a}{2}$.

The analogous relations  for arbitrary $n$
 are drastically simplified in the limit
when the space-time dimension $d\rightarrow\infty$. Of course,  the
real world
 corresponds to $d=4$, but we expect that the expansion $1/d^k$ works
well enough and errors are of order $1/d\simeq 25\%$.
Anyhow, we are not going to use this limit for
the numerical estimates. Instead, we want to use this limit
in order to understand
the general structure of the $\la \vec{k}_{\perp}^{2n} \ra $
(behavior at $n\rightarrow\infty$, in particular).
In this limit,  discarding the small terms related to
the creation of the additional $\bar{q}q$ pair\footnote{Indeed,
the additional terms which appear in this derivation have the form
$[\hat{iD},\sigma_{\mu\nu}
 G_{\mu\nu}]\sim(\Dm G_{\mu\nu})\gamma_{\nu}\sim
\bar{q}\gamma_{\nu}\lambda^aq\gamma_{\nu}$. It can be checked that
these terms are negligible small.},
we get the following simple formula for
 $\la \vec{k}_{\perp}^{2n} \ra_P$
  expressed exclusively in terms of condensate
 $\la \bar{q}(ig\sigma_{\mu\nu}
 G_{\mu\nu})^n q \ra$, which is the leading operator in
$d\rightarrow\infty$
limit:
 \be
\label{n1}
\la \vec{k}_{\perp}^{2n} \ra_P\simeq C^n
 \frac{\la \bar{q}(ig\sigma_{\mu\nu}
 G_{\mu\nu}^a\frac{\lambda^a}{2})^n q \ra }{\la \bar{q}q\ra},
{}~~~~ d\rightarrow\infty
\ee
with some constant $C$.

Now let us come back to
 the formula (\ref{7}). It is clear that the calculation of the mean
value of the quark transverse momentum in the pion and the
evaluation of some vacuum characteristics are two sides of the same coin.

Let us derive the same relations (\ref{7}) in a different way.
Consider for
this purpose the following correlator
\be
\label{9}
T_{\nu..\mu_n}=i\int dx e^{iqx}
\la 0|T\{\bar{d}i\gamma_5O_{\mu_1..\mu_n}
u(x), \bar{u}\gamma_{\nu}\gamma_5d(0)\}|0 \ra=
q_{\nu}T_{\{\mu_n\}}(q^2)+...
\ee
with  arbitrary local operator $O_{\mu_1..\mu_n}$. We
extracted the special kinematical structure $q_{\nu}T_{\{\mu_n\}}$ in
such  a way that the tenzor   $T_{\{\mu_n\}}$ does not depend on
$q_{\mu}$, but depends only on metric $g_{\mu_i\mu_j}$.
The leading contribution to $T_{\{\mu_n\}}$ at $q^2\rightarrow\infty$
can be easily calculated and is given by
\be
\label{10}
T_{\{\mu_n\}}(q^2)=\frac{-2}{q^2}
\la\bar{q}O_{\mu_1..\mu_n}q\ra +0(\frac{1}{q^4}).
\ee
On the other hand, the only   particle which can contribute
to this correlator is $\pi$ meson, because
$a_{\nu}=\bar{u}\gamma_{\nu}\gamma_5d $ is conserved
in the chiral limit
and matrix elements $\la 0|a_{\nu}|\pi'\ra$ for the massive
pseudoscalar particles are
proportional to $ \partial_{\nu}a_{\nu}\sim m_q\rightarrow 0 $.
At the same time, the axial mesons do not contribute to the
kinematical structure we chosen $q_{\nu}T_{\{\mu_n\}}$.

Thus, the whole result (\ref{10}) comes exclusively from the
$\pi$ meson   and any corrections to this statement are proportional
to the non-conservation of the current
$ \partial_{\nu}a_{\nu}\sim m_q  $ ;  in particular,
the next term proportional to $1/q^4$ comes with the small factor $m_q$.
Therefore, the correlation function $T_{\{\mu_n\}}$ looks
like a huge
peak related to $\pi$ meson contribution plus some
function on $q^2$ which is nontrivial, but
 proportional to the small number $m_q$.
Using the dispersion relation and retaining only $\pi$ meson
contribution to $Im T_{\{\mu_n\}}$ we recover the formula
(\ref{7}).
The great usefulness of this derivation will be appreciated soon,
in the analysis of the axial $wf$.
  \vskip .3cm
\noindent
{\bf 3. $\la \vec{k}_{\perp}^{2n} \ra_A$. Leading twist
 wave function.}
  
We define the pion
axial wave function $if_{\pi}q_{\mu}\phi_A=
\la 0|\bar{d}(z)\gamma_{\mu}\gamma_5  u(-z)|\pi(q)\ra $
and the corresponding
mean values of the quark transverse distribution
in the following way:
\be
\label{11}
\la 0|\bar{d}\gamma_{\nu}
\gamma_5 (\Dm t_{\mu})^{2n} u|\pi(q)\ra=if_{\pi}q_{\nu}
 (-t^2)^n\frac{(2n-1)!!}{(2n)!!}\la \vec{k}_{\perp}^{2n} \ra_A.
\ee
First of all let us show that the calculation of eq.
(\ref{11}) is reduced to the calculation of the
  matrix elements depending only on $G_{\mu\nu}$:
 We start from the simplest case and consider the following
matrix element \cite{Zhit2}:
 \be
\label{13}
\la 0|\bar{d}\gamma_{\nu}\gamma_5 \frac{1}{2}[\vec{iD}_{\mu_1}
\vec{iD}_{\mu_2}+\vec{iD}_{\mu_2}
\vec{iD}_{\mu_1}] u|\pi(q)\ra= \\  \nonumber
 if_{\pi}\{Ag_{\mu_1 \mu_2}q_{\nu}+B[g_{\mu_1 \nu}q_{\mu_2}+
g_{\nu \mu_2}q_{\mu_1}]\}
\ee
{}From the definition of $\la \vec{k}_{\perp}^{2} \ra_A$ it is clear,
that $\la \vec{k}_{\perp}^{2} \ra_A=-2A$. At the same time,
multiplying (\ref{13}) by metric tenzor
and using   equations of motion, the constant $A$ can be
expressed
in terms of the following matrix element:
\be
\label{14}
\la 0| \bar{d}\gamma_{\nu}\gamma_5(\frac{-ig}{2})
 \sigma_{\lambda\sigma}
 G_{\lambda\sigma} u|\pi(q)\ra=
\frac{18}
{5}if_{\pi}q_{\nu}A
\ee
 In order to express
 $\la \vec{k}_{\perp}^{4} \ra_A$ in the same terms,
 we consider the following
matrix element
\be
\label{15}
\la 0|\bar{d}\gamma_{\mu}\gamma_5\frac{1}{4!}\sum_{
\{\mu_1,\mu_2, \mu_3,\mu_4 \}}(\vec{iD}_{\mu_1
}\vec{iD}_{\mu_2} \vec{iD}_{\mu_3}\vec{iD}_{\mu_4})
 u|\pi(q)\ra.
\ee
By the reason mentioned above, we have to consider
on the right hand side of this equation the most general
kinematical structure which includes only the first
power of $q_{\mu} $ and which is symmetric
under $\mu_1,\mu_2, \mu_3,\mu_4 $ permutations; the rest
 indexes are cared by the
metric tenzor $g_{\mu\nu}$.
Multiplying (\ref{15}) by $ g_{\mu\nu} $, using
the equation of motion
and skipping all small terms which are related to
 creation
of the additional $\bar{q}q $ pair (for
justification, see below),
 we arrive to the following result:
\be
\label{16}
 iq_{\rho}f_{\pi} \la\vec{k}_{\perp}^{4}\ra_A=
  \frac{3}{8}\la 0| \bar{d}\gamma_{\rho}\gamma_5
(\frac{-ig}{2}
 \sigma_{\lambda\sigma}
 G_{\lambda\sigma})^2 u|\pi\ra+ ~~~~~~~~~~~~~~~~ \\ \nonumber
\frac{3}{16}
\la 0| \bar{d}\gamma_{\rho}\gamma_5g^2
 G_{\mu\nu}   G_{\mu\nu}u|\pi \ra
+\frac{1}{72}
\la 0| \bar{d}\gamma_{\mu}\gamma_5g^2
 (G_{\nu\rho }   G_{\mu\nu}+G_{\mu\nu }   G_{\nu\rho})u|\pi \ra .
\ee
Again, the problem is reduced to the calculation of
some matrix elements, which can be estimated by the method,
described in the previous section and which
we are going to use now.

In order to calculate the $\la \k\ra_A$ let us consider the
following correlator (compare with (\ref{9})):
\be
\label{17}
 i\int dx e^{iqx}
\la 0|T\{\bar{d}\gamma_{\mu}\gamma_5ig\sigma_{\lambda\sigma}
 G_{\lambda\sigma}
u(x), \bar{u} i\gamma_5d(0)\}|0 \ra=
q_{\mu}T (q^2)+...
\ee
In comparison with the previous case, the current
$J_{\mu}=\bar{d}\gamma_{\mu}\gamma_5ig\sigma_{\lambda\sigma}
 G_{\lambda\sigma}u $ is not conserved even in the chiral limit,
but it is``almost" conserved $\partial_{\mu}J_{\mu}
\sim(\Dm G_{\mu\nu})
\sim\bar{q}\gamma_{\nu}q$
in a sense, that the non-conservation is small and related to the
production of the additional $\bar{q}q$ pair.
{}From the point of view of the expansion $T(q^2)$ at large $q^2$
it means that the main contribution is given by
\be
\label{18}
T(q^2)=\frac{i}{q^2}\la\bar{q}ig\sigma_{\lambda\sigma}
 G_{\lambda\sigma}q\ra,
\ee
like in the previous case (\ref{10}); the corrections
to this formula
are suppressed by the loop factor $\frac{\alpha_s}
{\pi}\simeq 0.1$
. We neglect them in the following.
Using a dispersion relation and retaining the $\pi$ meson
contribution only \footnote{The spectral density $Im T(s)$
falls off
quickly at large $s$ in the chiral limit. This is an
additional justification
for the keeping only $\pi$ meson contribution.}, one gets
 \cite{Zhit2}:
\be
\label{19}
\la \vec{k}_{\perp}^{2} \ra_A=\frac{5}{36}\frac{\la
\bar{q}ig\sigma_{\mu\nu}
 G_{\mu\nu}^a\frac{\lambda^a}{2} q \ra }{\la \bar{q}q\ra}
\simeq \frac{5 m_0^2}{36}\simeq 0.1GeV^2.
\ee

The same procedure can be applied for any operator whose
matrix element we are interested in.
The result for $\la\vec{k}_{\perp}^{4}\ra_A$ is:
\be
\label{20}
  \la\vec{k}_{\perp}^{4}\ra_A=
 \frac{1}{8}\{   \frac{-3\la \bar{q}g^2\sigma_{\mu\nu}
 G_{\mu\nu} \sigma_{\lambda\sigma}
 G_{\lambda\sigma}q \ra}{4\la\bar{q}q\ra}
+ \frac{ 13\la \bar{q}g^2
 G_{\mu\nu}   G_{\mu\nu}q \ra}{ 9 \la\bar{q}q\ra}\},
\ee
The problem, like in the previous case is reduced to the
analysis of the mixed vacuum condensates.
 \vskip .3cm
\noindent
{\bf 4.Analysis of the condensates.}

Let me, first of all,  formulate the method  we follow
in this section
in order to estimate the higher dimensional condensates.
The idea is as follows. Consider the correlation function
$T_{\mu}=i \int dx e^{iqx}\la T\{J_{\mu}(x),J(0)\}\ra$ at
large
$q^2$. If the currents $J_{\mu},J$ are chosen in  such a
way, that
in the chiral limit the perturbative contribution is zero and the
current $J_{\mu}$ is ``almost" conserved (in a sense of the
 previous
section), we end up with the leading (at $q^2\rightarrow\infty$)
contribution
in the form $\frac{\la O \ra}{q^2}$, like (\ref{18}), plus
 some nontrivial
function, but with small coefficient, like $\frac{\alpha_s}
{\pi}, or ~~m_q$,
 in front of it. Besides that, if we know the $\pi$-meson
matrix elements
$\la 0|J_{\mu}|\pi\ra, \la 0|J |\pi\ra$ exactly, we can
find the
condensate $\la 0\ra$ by collecting all leading terms
proportional  to $1/q^2$.
 To convince the reader in correctness
of this method, we will  explicitly calculate
the loop corrections as well in the   example which follows.

In order to calculate the condensate
$\la \bar{q}g^2\sigma_{\mu\nu}
 G_{\mu\nu} \sigma_{\lambda\sigma}
 G_{\lambda\sigma}q \ra$,
let us consider the following
correlator:
\be
\label{21}
 i\int dx e^{iqx}
\la 0|T\{\bar{d}\gamma_{\rho}\gamma_5ig\sigma_{\lambda\sigma}
 G_{\lambda\sigma}
u(x), \bar{u} \gamma_5(ig\sigma_{\mu\nu} G_{\mu\nu})
d(0)\}|0 \ra=
q_{\rho}T (q^2)+...
\ee
The leading  contribution is determined by the following
 condensate, which
is unknown:
\be
\label{22}
T(q^2)=\frac{1}{q^2}\la\bar{q}(ig\sigma_{\lambda\sigma}
 G_{\lambda\sigma})^2 q\ra.
\ee
At the same time, both $\pi$ meson matrix elements
which enter to this correlator
are known -- one of them:
$\la 0|\bar{d}\gamma_{\rho}\gamma_5ig\sigma_{\lambda\sigma}
 G_{\lambda\sigma}
u(x)|\pi\ra$ is from (\ref{14})
and another one:
$\la 0|\bar{u} \gamma_5(ig\sigma_{\mu\nu} G_{\mu\nu})
d(0)\}|\pi \ra$ is from (\ref{4}) and can be found from PCAC.
Thus, we have finally the following very important equation:
\be
\label{23}
\la \bar{q}g^2\sigma_{\mu\nu}
 G_{\mu\nu} \sigma_{\lambda\sigma}
 G_{\lambda\sigma}q \ra=-\frac{
\la\bar{q}ig\sigma_{\lambda\sigma}
 G_{\lambda\sigma}q\ra^2}{\la\bar{q}q\ra}
\ee
Before we proceed
with the numerical estimates, let me briefly formulate the result of
the loop calculation for the correlator under consideration. After the
``borelization" procedure \cite{Shif1} the leading loop contribution
to the $T(M^2)$ is given by: $T(M^2)=-\frac{\alpha_s}{12\pi}M^2
\la\bar{q}q\ra$, where $M^2$ is Borel parameter which is determined by
the next power corrections (which have
 the same small factor $\alpha_s/\pi$)and usually
runs in the region $M^2\simeq 0.6\div 1 GeV^2 $.
This term is 30 times less than both
contributions: (\ref{22})  and  $\pi$ meson residue
(\ref{23}). This rate  gives us some intuition about
magnitude of the loop corrections   to the eq.(\ref{23}).
We neglect them in the rest of the paper.

The factorization prescription for the
condensate (\ref{22}) leads to the following formula:
\be
\label{24}
\la \bar{q}g^2\sigma_{\mu\nu}
 G_{\mu\nu} \sigma_{\lambda\sigma}
 G_{\lambda\sigma}q \ra =-\frac{K}{3}\la g^2G_{\mu\nu}^aG_{\mu\nu}^a\ra
\la \bar{q}q\ra,
 ~~K=\frac{3m_0^4}
{\la g^2G_{\mu\nu}^aG_{\mu\nu}^a\ra}\simeq 3,
\ee
where we have introduced the coefficient of nonfactorizability $K$
($K=1$ if the factorization would work).
The last equation follows from (\ref{23}) and actually demonstrates
that the factorization does not work.

Few comments are in order. First of all,
let us note that this result (the violation
of factorization by a factor 3) is in a full agreement
 with the (absolutely independent) analysis
 \cite{Zhit1} of the mixed
vacuum condensates of the dimension seven. The result
was based on the analysis of the heavy-light quark system and
supports the present consideration.
Besides that, let us note, that the VEV which appeared
in the analysis \cite{Zhit1} was actually some combination of
 condensates
$\la \bar{q}g^2\sigma_{\mu\nu}
 G_{\mu\nu} \sigma_{\lambda\sigma}
 G_{\lambda\sigma}q \ra$ and $\la \bar{q}g^2G_{\mu\nu}^aG_{\mu\nu}^a
 q\ra$. We found for that combination
more or less the same coefficient for the
nonfactorizability $K\simeq 3$. So, for the numerical estimates
it is natural to  assume that the factor $K\simeq 3$
is the universal one for all condensates of this kind.

As the second remark, we note that the phenomenon
  of non- factorizability
    for the mixed condensates  is not a big
surprise,   because we faced the analogous phenomenon early \cite{Zhit}
for the four-fermion condensates with an ``exotic" Lorentz structure.

As the last remark, and probably, the most important one,
 we want to emphasize that
the eq.(\ref{23}) can be considered as the recurrent relation
between $\la\bar{q}ig\sigma_{\lambda\sigma}
 G_{\lambda\sigma}q\ra$ and $\la \bar{q}g^2\sigma_{\mu\nu}
 G_{\mu\nu} \sigma_{\lambda\sigma}
 G_{\lambda\sigma}q \ra$ with the dimensional coefficient $m_0^2$.
We can repeat this procedure for arbitrary $n$ with result:
\be
\label{25}
\la\bar{q}(ig\sigma_{\lambda\sigma}
 G_{\lambda\sigma})^n q\ra=(m_0^2)^n\la \bar{q}q\ra,
\ee
which we expect is to be correct one with a high accuracy
$\sim\alpha_s/\pi$
\footnote{The analogous result for the
 $\la\bar{q}(  G_{\lambda\sigma})^n q\ra $ looks as follows:
$\la\bar{q}( G_{\lambda\sigma}^a
 G_{\lambda\sigma}^a)^n q\ra=(m_1^4)^n\la \bar{q}q\ra $, but with
unknown coefficient $m_1^4$.}.
 The formula (\ref{25}) is the {\bf main result} of this section.
First of all it demonstrates a very simple relation between condensates
and gives  very important phenomenological information for
modelekonstructors of the QCD vacuum.
 Secondly,  bearing in mind that these condensates
are directly related to the moments $\la \vec{k}_{\perp}^{2n} \ra $,
the formula (\ref{25}) gives  nontrivial information about
 nonperturbative
pion wave function $\psi(\k, x)$.
  \vskip .3cm
\noindent
{\bf 5.Constraints on the nonperturbative wave function $\psi_A(\k, x)$.}

Now we are going to consider some applications of the obtained results.
First of all let us consider the pseudoscalar $wf$ and its moments.
{}From eqs. (\ref{4},\ref{8}) we have the following ratio
for $ {\la \vec{k}_{\perp}^{4}\ra_P }/{\la \vec{k}_{\perp}^{2}\ra_P^2} $:
\be
\label{26}
\frac{\la \vec{k}_{\perp}^{4}\ra_P }{\la \vec{k}_{\perp}^{2}\ra_P^2}
= K~~\frac{8\la g^2G_{\mu\nu}^aG_{\mu\nu}^a\ra}{9m_0^4}\simeq
0.6K\simeq 2,
\ee
where we assumed   the universality for the nonfactorazibility
 factor   $ K\simeq 3 $, as explained above.   Let us
 remark, that     we know the condensate
(\ref{23}), which  enters  to the expression for
 $ \la \vec{k}_{\perp}^{4}\ra_P $
  ``almost" exactly. However, we do not possess such precise
information for the second
VEV which contributes to $ \la \vec{k}_{\perp}^{4}\ra_P $
with  the same weight. In spite of this fact
we expect a good accuracy for the
ratio, because both contributions go with the
{\it same} signs, and because some independent combination of these
VEVs leads to the same coefficient $K\simeq 3$, see above.
 From the pure theoretical point
 of view, by considering the
  limit $ d\rightarrow\infty $, we can argue that there is not any
cancellation
between different terms (most dangerous thing which could happen!).

This ratio  actually is  some expression  of the fluctuations of
the momentum.
The result (\ref{26}) demonstrates that the distribution function
for the pseudoscalar function is rather compact in the momentum
space in a big contrast with the analogous distribution for the
leading twist wave function.

For the axial $wf$  from the
formulae (\ref{19},\ref{20}) we have the following ratio
(instead of eq.(\ref{26})):
\be
\label{27}
\frac{\la \vec{k}_{\perp}^{4}\ra_A }{\la \vec{k}_{\perp}^{2}\ra_A^2}
\simeq 3K~~\frac{ \la g^2G_{\mu\nu}^aG_{\mu\nu}^a\ra}{m_0^4} \simeq  5\div 7,
\ee
which tells us that the fluctuations of the transverse momentum
are much larger for the axial $wf$ than for the pseudoscalar $wf$
\footnote{Let us note, that the same ratio was examined in the recent
preprint \cite{Miller} by the method of \cite{Cher2} with the result $
\frac{\la \vec{k}_{\perp}^{4}\ra_A }{\la \vec{k}_{\perp}^{2}\ra_A^2}
=9$ which  numerically is not far away from our estimates (\ref{27}).
 However from our opinion, the corresponding sum rules are not
suitable for the  calculations of $\la \vec{k}_{\perp}^{n}\ra_A$.
  We see at least few reasons  for that. First of all,
the corresponding spectral density increases very quickly $\sim s^n$
with $s$ and there is no reason for $\pi$ meson saturation. As a
consequence of it, there is a very strong dependence on $S_0$
which is far away ($S_0\simeq 0.4 GeV^2$) from the standard value
$S_0\simeq 1GeV^2$.  By some kinematical reasons, the 4-quark
condensate, which
used to be the most important one, does not contribute.
It clearly means, that the next power corrections, proportional
to $\la\bar{q}(ig\sigma_{\lambda\sigma}
 G_{\lambda\sigma})^n q\ra\la\bar{q} q\ra$ become very important.
Thus, our point is that the corresponding sum rules do not work
and moments $ \la \vec{k}_{\perp}^{2n}\ra$
 can not be extracted in such way.}.

It is interesting to note, that the analysis of the distribution
function for the longitudinal momentum $\la \xi^{2n}\ra$ \cite{Cher2}
shows   the same phenomenon:
deviation of longitudinal moments from their asymptotic values
is large for the axial $wf$ and very modest for the pseudoscalar $wf$.

In principle we are ready  to model  $\psi(\k, x)$.
However, as is known, the knowledge of a finite number of moments is
not sufficient to completely determine the $\psi(\k, x)$.
The behavior of the
asymptotically distant terms is  a very important thing as well.
Let me remind you of an example of   the well-known case for
the longitudinal
distribution   which determined by the  $\phi( \xi)$.
If $\la\xi^m\ra$ would
behave as $\sim 1/m$ at large $m$, it would
mean  that the integrand
 $\int_{-1}^1d\xi \xi^m\phi(\xi)\sim 1/m$ implying that
$\phi(\xi\rightarrow
\pm 1)\rightarrow constant.$
At the same time, the very likely assumption that the $\pi$-meson fills
a finite duality interval in the dispersion relation at any $m$
means that $\int_{-1}^1d\xi \xi^m\phi(\xi)\sim 1/m^2$ \cite {Cher}.
It  unambiguously  implies the behavior \\
$\bullet {\bf 1}~~~~~~~~ \phi(\xi\rightarrow
\pm 1)\rightarrow (1-\xi^2)$.\\
A simple model wave function which
meets this requirement as well as possesses a large value of the
 lowest moment, $\la\xi^2\ra\simeq 0.4$ has been proposed in ref.
\cite{Cher2}:
\be
\label{28}
 \phi(\xi)=\frac{15}{4}(1-\xi^2)\xi^2,~~~\la\xi^0\ra=1,
{}~~~\la\xi^2\ra=0.43
\ee
We want to emphasize that the ``endpoint" behavior
(which corresponds to asymptotically distant terms) was crucial
in this analysis.

In terms of $\psi(\k, \xi=2x-1)$ this
constraints can be rewritten in the following form:
 $\int d\k \psi(\k, \xi\rightarrow\pm 1)\sim (1-\xi^2) $.
 The analogous assumption, that the $\pi$-meson fills
a finite duality interval in the corresponding dispersion
 relation at any $n$,
where $n$  is  related to $n$- transverse moment
$\la \vec{k}_{\perp}^{2n} \ra $
gives the following constraint:
 \be
\bullet {\bf 2}~~~~~~~~~~~~~~~~~~~~~~~
\int d\k
\vec{k}_{\perp}^{2n}\psi(\k, \xi\rightarrow\pm 1)\sim (1-\xi^2)^{n+1}
{}~~~~~~~~~~~~~~~~~~~.\nonumber
   \ee
This constraint is extremely important and implies that the $\k$
dependence of the   $\psi(\k, \xi=2x-1)$ comes
{\bf exclusively in the combination}
$ \k/(1-\xi^2)$ at $\xi\rightarrow\pm 1$.  The byproduct of this
constraint can be formulated as follows. The standard assumption
on factorizability of the $\psi(\k, \xi ) =\psi(\k )\phi(\xi)$
{\bf does contradict} to the very general property of the theory
formulated above.

The next constraints come from the calculation $\la\k\ra_A$,
(\ref{19})
and $\la\vec{k}_{\perp}^{4}\ra $, (\ref{27}):
\be
\bullet{\bf 3}
\int d \k \int_{-1}^1d \xi \psi(\k,\xi)=1,
\la\k\ra=
 \frac{5m_0^2}{36}\simeq
0.1 GeV^2,~~
\la\vec{k}_{\perp}^{4}\ra
 \gg \la\k\ra^2.~~~~~ \nonumber
\ee
These constraints actually give the general scale of the
$\psi(\k,\xi)$. Besides that, the large fluctuations of
 the momentum
mean, first of all, that the distribution of the transverse
quark momentum
is wide and it is not concentrated about $\k\simeq 0$.

Our last constraint comes from the analysis of the
asymptotically distant
terms. We do not know   the behavior of the asymptotically
large moments $\la\vec{k}_{\perp}^{2n}\ra $ at
$n\rightarrow\infty$ exactly.
But we do know
 the dependence on $n$ in the limit when the space-time dimension
$d\rightarrow\infty$, (\ref{n1})\footnote{We derived this formula
for pseudoscalar function only, but absolutely the same formula
takes place for the axial $wf$ as well. The derivation is the same.}.
 We expect (and this is the main assumption) that
in the real world the functional dependence will  not  be changed.
Thus, for the  asymptotically large $n$ we expect the following
behavior:
 \be
\bullet{\bf 4}~~~~~~~~~~~~~~~~~~~~~~~~
\la \vec{k}_{\perp}^{2n} \ra\Rightarrow
 \frac{\la \bar{q}(ig\sigma_{\mu\nu}
 G_{\mu\nu}^a\frac{\lambda^a}{2})^n q \ra }{\la \bar{q}q\ra}
\sim ( m_0^2 )^n, ~~~~~~~~~~~~~\nonumber
 \ee
where at the last stage  we have  used  the eq.(\ref{25}), which
is suppose to be valid for the arbitrary $n$.
It is very important that the right hand side of this equation is
constant  and not zero. We are not going to use this
 constant
in our model as input parameter; the only fact we need,
that this is not zero. Let us note, that any mild function on $n$, like
$1/n$ or even $\exp(n)$ on the right hand side can not be ruled out.
The only effect it brings, is some rescale of the dimensional
parameter:
$m_0^2\rightarrow\tilde{m_0^2}=\sqrt[n]{n}m_0^2=m_0^2 $ or
$m_0^2\rightarrow\tilde{m_0^2}=e m_0^2  $, which anyhow, is unknown.
The same constraint can be obtained from the dispersion  relation
with the  assumption that the $\pi$ meson fills a finite
duality interval $S_0$ at any $n$, like
in the analysis ($\bullet {\bf 1}$). In this case, instead of $
(\bullet{\bf 4}) $ we get
\be
f_{\pi}^2\la \vec{k}_{\perp}^{2n} \ra \Rightarrow
\frac{3S_0^{n+1}(2n+2)!!}{8 \pi^2(2n+3)!!}\sim \frac{S_0^{n+1}}{n},\nonumber
\ee
which can be reduced to the previous one.

Very important consequence of this constraint
can be easily seen if we rewrite it in the following form:
\be
\bullet {\bf 4}~~~~~~~~~~~~~~~~~~~~~~~~~~~~~~~~~~~~
 \int d \k   \psi(\frac{\k}{1-\xi^2},\xi)
(  \vec{k}_{\perp}^{2} )^n
\Rightarrow 1,~~~~~~~~~~~~  \nonumber
\ee
which means that the $\psi(\frac{\k}{1-\xi^2}) \Rightarrow
\delta(\frac{\k}{1-\xi^2}-S)$ for sufficiently large $\k$.
We have to pause here in order to explain the   general idea of
the Wilson operator expansion (OPE) in the given context  and the
term ``sufficiently large $\k$" in particular.

As is known, all VEVs within OPE are defined in such a way, that
all gluon's and quark's
 virtualities, smaller than some parameter $\mu$
(point of normalization)
are hidden in the definition of the ``nonperturbative vacuum matrix
elements". All virtualities larger than that should be taken into
account perturbatively, see \cite{Shif2} for detail discussion of
this problem. At the same time, from the exact PCAC relations, like
(\ref{4}) or(\ref{8}) it is clear that moments of the wave functions
are related to condensates, defined as explained above. Thus, all
transverse moments are defined in the same way as condensates do.
This is, actually, the {\bf definition of the nonperturbative
wave function $ \psi( \k, \xi)$}, through its moments
which can be expressed
in terms of nonperturbative vacuum condensates.

Now it clear, what   we mean by term ``sufficiently large $\k$".
By that we mean the largest virtuality which have been taken into
account
in construction of $ \psi( \k, \xi)$, or (what is the same)
in the definition of condensates. The corresponding dimensional
parameter can be expressed in terms of some condensate (see bellow)
and it
depends on $\mu$ only logarithmically, like condensates do.
 The important consequence of the definition can be
formulated in the following way: The nonperturbative $wf$
even for sufficiently large $\k$ does not behave like
$1/\k$ as is frequently assumed.
 In order to understand this statement let us imagine
that we calculate the gluon condensate $\la G_{\mu\nu}^2\ra$.
It is clear that we will not substitute the gluon propagator
in the form $1/k^2$ in this calculation, because, by
definition, the perturbative contribution should be
subtracted in the calculation of the nonperturbative condensate.
Have in mind that our definition of the nonperturbative $wf$
is formulated in terms of the vacuum condensates, it is clear
that the same statement is concerned to the
nonperturbative wave function as well and its behavior
has nothing to do with $1/\k$.

We want to model the simplest version of the wave function which meets all
requirements formulated above. First of all, as was explained,
the $\k$ dependence comes only in the combination
$\psi(\frac{\k}{1-\xi^2})$ in order to meet the
requirement ($\bullet {\bf 2}$). In order to satisfy the
constraint ($\bullet {\bf 4}$), we have to assume that at
sufficiently large $\k$ we have  a $\delta(\frac{\k}{1-\xi^2}-S)$
-like function,
whose moments correctly reproduce the constant on the right hand side
of the   equation; the    $S$ is some input dimensional parameter,
which will be expressed in terms of $\la\k\ra$.

 Our next step is to satisfy the constraint ($\bullet {\bf3}$).
It is clear that the $\delta$ function proposed above does not
provide  a noticeable fluctuations of the momentum
 In order to meet this requirement, we have to spread out
the distribution function
between $\k\sim0$ and $\k\sim S$ in such a way, that the overall area
will be the same, but moments should satisfy to this requirement.
  In principle, it can be done in arbitrary way. The simplest way
to make the $wf$ wider is to put another $\delta$ function
at $\k=0$.
 With these remarks in mind we propose the following "two-hump"
({\it again!})
nonperturbative wave function which meets all requirements
discussed above:
\be
\label{m}
\psi(\k,\xi)=[A\delta(\frac{\k}{1-\xi^2}-S)+B\delta(\frac{\k}{1-\xi^2})]
[g(\xi^2-\frac{1}{5})+\frac{1}{5}]   \nonumber
\ee
\be
\int d \k \int_{-1}^1d \xi \psi(\k,\xi)=1,
{}~~~~~~~~ A+B=\frac{15}{4}
\ee
\be
\phi(\xi)\equiv \int d \k \psi(\k,\xi)=\frac{15}{4}(1-\xi^2)
[g(\xi^2-\frac{1}{5})+\frac{1}{5}]
\nonumber
  \ee
\be
\la\vec{k}_{\perp}^{4}\ra
 \simeq 5\la\k\ra^2 \Rightarrow A=7/8,~~~ g=1,~~~
\la\k\ra=
 \frac{5m_0^2}{36}\Rightarrow S=\frac{15}{2}\la\k\ra\simeq 0.8 GeV^2
\nonumber
\ee
Few comments are in order. We put the common factor
 $[g(\xi^2-\frac{1}{5})+\frac{1}{5}] $ in the
front of the formula  in order to reproduce  the light
cone   $\phi(\xi)$ function with arbitrary $\la\xi^2\ra$.
For $g=0$ it corresponds to the asymptotic $wf : \phi(\xi)
=3/4 (1-\xi^2)$. For   $g=1$ we reproduce the $\phi(\xi)_{CZ}=
15/4 (1-\xi^2)\xi^2$ with $\la \xi^2\ra\simeq 0.43$.
We still believe in large value for the moment $\la \xi^2\ra $
  in spite of the criticism from
the  ref.\cite{Mikh}, who found much less value for $\la\xi^2\ra$.
This is not the place to discuss this question in   more detail,
but I want to make a comment that there are few questions
to be answered ( theoretical, as well as phenomenological ones)
before the approach advocated by authors of ref.\cite{Mikh}
can be considered as is well defined and based on solid ground.
   At  the same time, the people who prefer to use a smaller value for
$\la\xi^2\ra$, can make   their own choice by changing
 the parameter $g$
 in the eq.(\ref{m}).
Let me remind the relation $ \la\xi^2\ra=1/5+8g/35$ between
$\la\xi^2\ra$ and parameter $g$ from the formula (\ref{m})
 for doing so.
It will
 not spoil the constraints discussed above.

Let me emphasize, that the formula (\ref{m}) is not destined
for the
precise  fitting
of the experimental data and it is given as illustration only.
By the physical reasons it is clear that the $\delta$ functions should be
spread out in  some way. This procedure is definitely model dependent.
 We only want to attract attention
on the asymmetric
form of the $wf$ in the  $\vec{k}_{\perp}$   space. This
qualitative property of the $wf$ comes from the large magnitude
of the ratio $ r=\la\vec{k}_{\perp}^{4}\ra /
   \la\k\ra^2$.

As the last remark, let me point out, that the description of
exclusive reactions at experimentally accessible momentum
transfers ($ 1 GeV^2< Q^2<30 GeV^2 $) and the problem
of extracting some information about nonperturbative $wf$
are two different problems. It is was advocated
by       Radyushkin  and collaborators \cite{Ditt}, \cite{Rad1}
 and Isgur,  Llewellyn Smith \cite{Isgur}
for a long time that the asymptotically leading contributions
do not describe the experimental data at moderate $Q^2$, for recent
review of this subject, see \cite{Rady}.
Some confirmation of this statement came recently from
\cite{Kroll}, where it was shown that the inclusion
of the intrinsic $\k$- dependence and Sudakov suppression
leads to the self-consistent calculation, but the
obtained magnitude is too
small with respect to the data even if $\phi_{CZ}$ asymmetric
function is used.
It is very likely, that the "soft" contributions play an
important role in this region.

It turns out, that the very unusual shape of the $\psi(\k,\xi)$
described above  supports
this idea and can {\bf imitate} the behavior of the
asymptotically leading contribution
at the intermediate momentum transfer.
Such a mechanism, if it is correct, would be an explanation
of the phenomenological success of the dimensional
counting rules at a very modest $Q^2$.
 We are going to discuss this question in  more detail somewhere else
\cite{Zhit4}.

\noindent
{\bf 6. Conclusion.}

Let me formulate the main results of this paper.

First of all, we used the standard definition of the nonperturbative
$wf$  through its moments and we expressed these moments in terms of the
vacuum condensates. From this construction it becomes clear that
the definition of the vacuum condensates (within OPE), with subtraction
of the perturbative contributions,
 and the definition of the nonperturbative
$wf$ is one and the same problem.

Secondly, we formulated some constraints on the nonperturbative $wf$;
they have very general origin, and thus, they should be satisfied
for any model. There is no reason to repeat these constraints
($\bullet {\bf 1}-\bullet {\bf 4}$) from the previous section again;
the only comment we want to make here is the following: the standard
assumption on factorizability of the $wf:~ \psi(\k, x ) =\psi(\k )\phi(x)$
can not be matched with these constraints.

The last result which deserves to be mentioned here and which probably
is interesting by its own, regardless to the question of nonperturbative
$wf$, looks as follows
$\la\bar{q}(ig\sigma_{\lambda\sigma}
 G_{\lambda\sigma})^n q\ra=(m_0^2)^n\la \bar{q}q\ra $.
 I am not aware of any model of the QCD vacuum, which satisfies to this
``almost" exact relation between mixed vacuum condensates.

\noindent
{\bf 7. Acknowledgements.}
 
I thank Tolya Radyushkin  for useful comments,
insights and  for the pointing  out that my previous
definition  of the moments did not correspond to the standard
definition of the distribution function, though
the numerical difference is not large (about 25$\%$)\footnote
 { The correspondence
is reached only in the limit $d\rightarrow\infty$}.

This work is supported by the Texas National Research
Laboratory Commission under  grant \# RCFY 93-229.

\end{document}